\documentclass[manuscript,screen,authorversion]{acmart}
\AtBeginDocument{%
  \providecommand\BibTeX{{%
    \normalfont B\kern-0.5em{\scshape i\kern-0.25em b}\kern-0.8em\TeX}}}

\setcopyright{rightsretained}
\copyrightyear{2023}
\acmYear{2023}

\usepackage{bbm}
\usepackage{makecell}
\usepackage{subcaption}
\usepackage{adjustbox}
\usepackage{placeins}

\begin{document}

\title{A Justice-Based Framework for the Analysis of Algorithmic Fairness-Utility Trade-Offs}

\author{Corinna Hertweck}
\authornote{Equal contribution.}
\email{corinna.hertweck@zhaw.ch}
\orcid{0000-0002-7639-2771}
\affiliation{%
  \institution{Zurich University of Applied Sciences, University of Zurich}
  \country{Switzerland}
}

\author{Joachim Baumann}
\authornotemark[1]
\email{baumann@ifi.uzh.ch}
\orcid{0000-0003-2019-4829}
\affiliation{%
  \institution{Zurich University of Applied Sciences, University of Zurich}
  \country{Switzerland}
}

\author{Michele Loi}
\authornotemark[1]
\email{michele.loi@polimi.it}
\orcid{0000-0002-7053-4724}
\affiliation{%
  \institution{Polytechnic University of Milan}
  \country{Italy}
}

\author{Eleonora Vigan\`o}
\email{eleonora.vigano@uzh.ch}
\orcid{0000-0002-1640-2763}
\affiliation{%
  \institution{University of Zurich}
  \country{Switzerland}
}

\author{Christoph Heitz}
\email{christoph.heitz@zhaw.ch}
\orcid{0000-0002-6683-4150}
\affiliation{%
  \institution{Zurich University of Applied Sciences}
  \country{Switzerland}
}

\renewcommand{\shortauthors}{Hertweck and Baumann, et al.}

\begin{abstract}
In prediction-based decision-making systems, different perspectives can be at odds: The short-term business goals of the decision makers are often in conflict with the decision subjects' wish to be treated fairly. Balancing these two perspectives is a question of values. However, these values are often hidden in the technicalities of the implementation of the decision-making system. In this paper, we propose a framework to make these value-laden choices clearly visible. We focus on a setting in which we want to find decision rules that balance the perspective of the decision maker and of the decision subjects. We provide an approach to formalize both perspectives, i.e., to assess the utility of the decision maker and the fairness towards the decision subjects. In both cases, the idea is to elicit values from decision makers and decision subjects that are then turned into something measurable. For the fairness evaluation, we build on well-known theories of distributive justice and on the algorithmic literature to ask what a fair distribution of utility (or welfare) looks like. This allows us to derive a fairness score that we then compare to the decision maker's utility. As we focus on a setting in which we are given a trained model and have to choose a decision rule, we use the concept of Pareto efficiency to compare decision rules. Our proposed framework can both guide the implementation of a decision-making system and help with audits, as it allows us to resurface the values implemented in a decision-making system.
\end{abstract}


\maketitle

\section{Introduction}

The increasing use of prediction-based decision-making systems has shown that this can easily lead to disadvantages for marginalized groups (see, e.g., \cite{machine-bias, amazon-hiring, br-retorio, obermeyer2019dissecting, buolamwini2018gender, eubanks2018automating}). These systems are very unlikely to achieve fairness because they are optimized for goals \textit{other than fairness}. Our framing hypothesis is that, besides pursing the decision maker's goal (e.g., to be as efficient or profitable as possible), a decision-making process should be fair towards the decision subjects, i.e., towards the individuals affected by the decisions.
Often, these two goals conflict \cite{kearns2019ethical}.\footnote{Note that they are not always in conflict as \citet{feder2021emergent} point out. If the primary goal of a decision maker is to achieve fairness, then the first five value-laden questions of our framework are still relevant, but the sixth one is not of any interest.} Navigating this trade-off requires making the values of the decision maker and the decision subjects explicit --- to the point where they can be expressed as mathematical formulas.
The perspective of fairness has been discussed in both computer science, coming up with many different so-called "fairness metrics" \cite{narayanan2018translation, mitchell2021algorithmic} and, for a much longer time, in philosophy. Philosophers have attempted to characterize ideally just institutions in terms of the set of moral principles they satisfy \cite{rawls1999theory} or to characterize those, and only those, inequalities that are ultimately morally important for justice \cite{sen2009idea}, thus providing a normative grounding to judgments about injustice reduction. So far, these two debates have mostly been developed apart from each other. However, the philosophical moral grounding problem is relevant for computer science since the criteria of fairness discussed within that discipline cannot be simultaneously fulfilled  \cite{kleinberg2016inherent, chouldechova2017fair, fairmlbook}.

It is important to note that the debate about appropriate fairness metrics is not a mathematical debate \cite{selbst2019fairness, wong2020democratizing}. As \citet{jacobs2021measurement} points out, the plethora of fairness definitions and the conflicts between them stem from the conflicting theories of fairness that they operationalize and reflect different values. Thus, it is a debate about values \cite{jacobs2021measurement} and one's beliefs about the world \cite{friedler2016possibility}.
Recent works \cite{wong2020democratizing, haeri2022promises} have highlighted the need for a deliberative process to explicate these values. \citet{wong2020democratizing} argues that the choice of fairness metric(s) is a choice of values and thereby inherently political. Consequently, \cite{wong2020democratizing} demands a democratization of this choice.

Here we propose a framework for eliciting and implementing moral values relevant to the choice of a fairness goal achievable by prediction-based decision-making. Our proposed framework elicits these values from decision makers and decision subjects through six value-laden questions. It also provides a simple way to set parameters of a prediction-based decision-making system such that it aligns with the agreed-on values. We assume a binary decision-making system where individuals are assigned probabilities, e.g., the predicted probability of repaying a loan. A decision rule takes this predicted probability as an input and makes the final binary decision. We also assume that it is possible to compare the consequences of these decisions for two socio-demographic groups (a privileged one and a disadvantaged one) in terms of the utility they generate for the decision subjects.

\begin{figure}[ht]
    \centering
    \includegraphics[width=\textwidth]{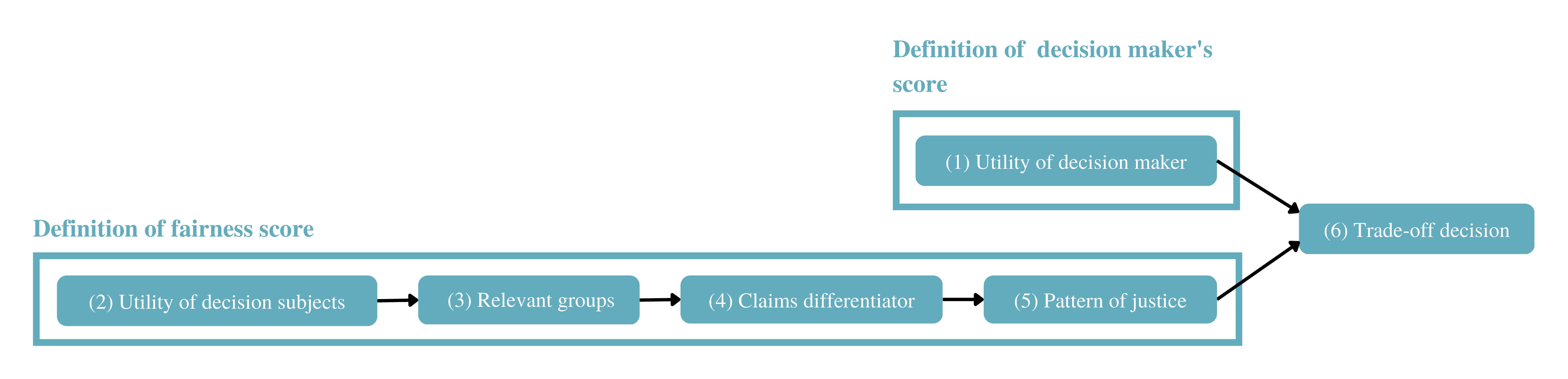}
    \caption{The six steps of our framework and their connections.}
    \label{fig:approach}
\end{figure}

The central idea of the framework is to specify one's normative preferences regarding six value-laden questions:
\begin{enumerate}
    \item \textbf{Utility of the decision maker:} How should we assess the benefit/harm that the decision maker derives from the decisions?
    \item \textbf{Utility of the decision subjects:} How should we assess the benefit/harm that the decision subjects derive from the decisions?
    \item \textbf{Relevant groups:} What groups of people are affected unequally by decision-making systems because being a member of a group is a (direct or indirect) cause of inequality?
    \item \textbf{Claim differentiator:} By virtue of which features can individuals morally demand equal consideration by the decision maker?
    \item \textbf{Pattern of justice:} Should the goal of justice be equality or some other distribution (e.g., maximizing the expectations of the worst-off group)?
    \item \textbf{Trade-off decision:} How strongly should fairness be pursued if it comes into conflict with the utility of the decision maker?
\end{enumerate}
As can be seen in Figure~\ref{fig:approach}, question (1) helps to derive a score of the decision maker's utility. Questions (2)-(5) allow us to define a morally appropriate fairness criterion and a score that expresses to what degree it is fulfilled. Question (6) then balances these two scores through a Pareto front that compares different possible decision rules. This step thus makes the trade-off explicit.

The rest of the paper is structured as follows: First, we highlight related work in Section \ref{sec:related-work}. In Section \ref{sec:setting}, we describe the general setting of prediction-based decision-making systems including two conflicting perspectives: the decision maker's and the decision subjects'. In this context, we explain the first value-laden choice of our framework. In addition, we introduce the notation that we will use throughout the paper. In Section \ref{sec:distributive-justice}, we describe a common structure of theories of justice that is represented in steps (2)-(5) of our framework. In Section \ref{sec:pareto-front}, we will present our suggestion for navigating the trade-off between the decision maker's goals and the fairness towards the decision subjects based on \cite{kearns2019ethical} --- the final step of our framework. We will then exemplify the value-laden choices in a case study in Section \ref{sec:examples}.
Finally, we discuss the limitations and merits of our framework in Section \ref{sec:limitations} and conclude in Section \ref{sec:conclusion}.

\section{Related Work}\label{sec:related-work}
Our paper combines several distinct developments in the recent literature on fairness in machine learning. The goal is to use them in a novel way to define an ethical framework for supporting the implementation of a fairer decision-making system. 

\paragraph{Utility-based view of fairness}
The first development is framing the problem of fairness in machine learning as the moral problem of justifying the distributive implications of the decisions based on those predictions. As not every individual derives the same benefit or harm from the same decision, a line of research has developed that formalizes the \textit{utility} or \textit{welfare} implications of such decisions for the individuals affected by them (in our framework, "decision subjects"). As pointed out by philosophers in the debate on whether the metric of justice should be resources, utility, or capabilities \cite{sen1980equality, lundgard2020measuring}, a resource (e.g., a loan offered by a bank) can be converted into different utility or capability levels by different types of individuals (e.g., those who are able vs. not able to repay a loan).

In the history of fair ML debate, this approach has been pioneered by several papers at the intersection of economics, political philosophy, and machine learning. \citet{heidari2018fairness} have proposed welfare-based definitions of fairness that take the effects of decisions into account and can be used as learning constraints.
\citet{heidari2019moral} have highlighted that what a fair distribution of utility looks like is influenced by one's claim to utility --- by which we mean the moral consideration that counts as a justification of inequality. In mapping the philosophical theory of equality of opportunity to group fairness metrics, they consider individual effort to be relevant for the moral claim one has to utility. This argument has been further developed in \cite{loi2021fair,baumann2022SDS_fairness_principle}. \citet{hertweck2021moral} and \citet{hu2020fair} showed, through philosophical arguments and with an empirical study (respectively), that enforcing fairness criteria can actually harm marginalized groups if it is not guided by utility considerations. 

Drawing on this literature, our framework views fairness in machine learning as a special case of the problem of selecting a distributive mechanism that is reasonably expected to achieve justice. In this view, the algorithm plays an analogous role as social institutions in traditional theories of justice, as proposed by \citet{heidari2019moral}.

\paragraph{Choice between conflicting fairness criteria}
The second development in the literature is the emergence of multiple, conflicting fairness criteria, lacking a systematic framework to choose among them. These criteria can be categorized in different ways.
Our approach can be seen as an extension of previous attempts to systematize the choice between \textit{group fairness criteria}.
Standard group fairness criteria compare metrics derived from the confusion matrix (such as true positive rates, false positive rates, positive predictive value, etc.) for two or more socio-demographic groups. Commonly, group fairness criteria demand equality in this comparison metric~\cite{binns2020apparent}. \citet{saleiro2018aequitas} and \citet{makhlouf2021applicability} provide a flow chart to guide the choice between these standard fairness criteria.

We share the goal of these papers: to create a framework for --- or even better, a morally principled solution to --- the problem of selecting a pertinent fairness criterion. In some respect, our approach can be regarded as an extension (and generalization) of \citet{heidari2019moral}, who also resolve the apparent conflict between fairness metrics by analyzing fairness at a higher level of abstraction: Appropriate fairness metrics can be directly derived from one's values. The novelty of our approach is to redefine the scope of the question: our problem is no longer to select between the items of an already given list of group fairness criteria (derived from the confusion matrix) but to formalize stakeholder values related to justice in such a way that a pertinent group fairness criterion will be determined. The group fairness criterion selected in this way can, but does not have to, correspond to one of those that can be specified by reference to the confusion matrix.\footnote{This is explained in detail in~\cite{Baumann2023unification}, where it is also shown that standard group fairness metrics --- i.e., (conditional) statistical parity, equalized odds, equality of opportunity, FPR parity, sufficiency, predictive parity, and FOR parity --- can be derived for specific choices of steps (2)-(5) of our framework.}

\paragraph{Relation to individual fairness and causality-based fairness}
Group fairness criteria are traditionally seen as opposed to individual fairness \cite{dwork2012fairness} or causal definitions of fairness \cite{kusner2017counterfactual}.%
\footnote{However, see \cite{binns2020apparent} for the claim that fair ML categories represent such distinctions as being sharper than they actually are.}
We note, \textit{inter alia}, that our paper addresses the two standard objections against group fairness criteria raised by these two alternative approaches: fairness gerrymandering \cite{dwork2012fairness} and causal irrelevance \cite{kusner2017counterfactual}. Philosophers \citet{hedden2021statistical} and \citet{long2021fairness} argued against classification parity (a specific statistical group fairness measure) by pointing to examples in which its violation was morally irrelevant. Remarkably, in both of these examples, the group variable is causally irrelevant by construction. Thus, these specific arguments against classification parity provide further support to the (much broader) causal irrelevance objection.
We shall later discuss how our approach responds to the criticism in question.

\paragraph{Fairness-utility trade-offs}
Finally, our framework brings together the aforementioned debate with proposals to evaluate trade-offs between the goal of the decision maker and fairness \cite{kearns2019ethical}.
One option to balance these two goals is to train a model on an objective function that combines the decision maker's utility and a fairness score as seen in \cite{kamishima2011fairness}. However, this option requires weighing the importance of both perspectives which can be difficult without having a clear idea of how big the trade-off may be.
Several works (e.g., \cite{corbett2017algorithmic, hardt2016equality, menon2018cost, baumann2022sufficiency}) have highlighted these conflicts and tried to quantify the trade-off between the two goals.
However, these works demonstrate the trade-offs for specific instantiations of the decision maker's goal (such as accuracy) and fairness (standard group fairness criteria).
In practice, we cannot assume that these specific formalizations represent the moral values of the decision makers and decision subjects in a given context.
As \citet{kearns2019ethical} highlight, the first step to balancing these two perspectives is therefore to make our values explicit. These values should guide how we formalize the decision maker's goal and fairness. The framework we propose provides a simple approach that builds on \cite{kearns2019ethical} to support the question of how to balance this definition of fairness with the utility of the decision maker.

\paragraph{Research gap}
What is still missing is a unified framework that guides stakeholders to identify the type of fairness goal they want to achieve (assisted by a menu of standard normative choices from political philosophy) and the degree to which they want to achieve the goal in conflict with the main purpose of the prediction algorithm.

\section{Prediction-Based Decision-Making}\label{sec:setting}

Our approach models fairness for prediction-based decision-making systems with distributive implications (i.e., purely predictive mechanisms relevant to epistemic views of justice \cite{fricker2007epistemic, anderson2012epistemic} fall outside the scope of this approach).
The goal of these systems is to make a decision $D$ based on a set of variables. Predictions are needed because the central variable that the decision is based on is not known at the time of decision --- we refer to this as the decision-relevant attribute $Y$.
In recruiting, for example, it is unclear whether an applicant will perform well; in medical applications, it is unclear whether a treatment will actually cure the patient. 
For the purpose of simplification, we assume that $D$ and $Y$ are binary: $D,Y \in \{0,1\}$.
The output of the predictor for a person with the attributes $X$ is a probability score $p=P(Y=1|X)$, which is used in the decision-making process.
A decision rule $r$ is a function that, for every individual, takes $p$ (and possibly other attributes) as an input and gives a decision as an output, e.g., “give a loan to everyone who has an estimated repayment probability of more than 80\%.”

In prediction-based decision-making systems, the decision maker typically makes many decisions of the same type.
Here we shall assume that the decision maker pursues reasonable goals and that there exists a metric that expresses the degree to which these goals have been achieved. We refer to this metric as the "decision maker's utility". In prediction-based decision-making systems, the decision maker typically makes many decisions of the same type. Their possible consequences can be identified and modeled probabilistically. Thus, the degree to which the decision maker's goal is achieved can be measured as \emph{expected utility} (utility weighted by probability). We assume that utility in this sense is something that decision-makers typically want to maximize.%
\footnote{
Depending on the context, there might be different boundary conditions, such as resource constraints, legal obligations, or business strategies.
}
This requires a \textbf{first value-laden choice}: how does one represent and measure the utility the decision maker wants to achieve through a given set of decisions?\footnote{The reason why we call this choice ``value-laden'' is that often it is impossible to derive a proper measure of utility in the sense we specified by simply observing the behaviors of decision-makers. In particular, in complex organizations, morally significant choices (such as in human resources) often pursue several goals simultaneously. The definition of a goal (even when the goal is defined as a weighted function of a plurality of goals) always involves a drastic simplification from the observed social reality, which can hardly be achieved without relying on some normative assumption.}

However, if the fairness of the decisions for the affected individuals should also be considered, the decision maker is required to deviate from their optimal decision rule, as this usually does not satisfy any social desideratum that is unequal to the decision maker's immediate goal (which is measured by the utility function).
This requires assessing the decision subjects' utilities for a given set of decisions to specify a morally appropriate definition of fairness --- constituting additional value-laden choices, which will be introduced in the following section.

\section{The Components of Fairness Metrics for Decision-Making}\label{sec:distributive-justice}

While attempting to achieve the goals for the decision maker, any prediction-based decision system relevant to our analysis coincidentally (and in some cases, unintentionally) distributes benefits (or harms) among members of society. We understand a fair prediction-based decision system to involve predictions and decision rules that combined can be reasonably expected to achieve a just distribution of benefits and harms across different groups. We turn to theories of justice in the tradition of political philosophy in order to determine what is a \textit{just} distribution. 

We characterize theories of justice by their answers to the following questions, which represent the \textbf{next value-laden choices}: What is, ultimately, distributed? Between whom is it distributed? Which subgroups should be compared? And how should it be distributed?~\cite{rawls1999theory, sen1980equality}.

\subsection{Utility of the decision subjects}\label{sec:utility-ds}
\textit{What is, ultimately, distributed?}

We will refer to what is being distributed, which could be positive in the case of a benefit and negative in the case of harm, as the \textit{utility of the decision subjects}. This builds on the line of welfare-based definitions of fairness described in Section \ref{sec:related-work}. Utility can be defined in different ways. We define well-being as what people have reasons to desire --- an "objective list" or "informed-desire" approach \cite{Griffin1986WellBeing,sep-well-being} and delegate the choice of a measure of utility to the hypothetical stakeholders that would employ this framework to arrive at their favored definition. Negative utility can be defined as what people desire \textit{not} to have.

\begin{definition}[Decision subject utility]
Decision subject utility is the amount of benefit or harm derived from receiving a certain decision. It is what people have (objective) reasons to desire.
\end{definition}

In our framework, we do not consider the overall level of utility of decision subjects but only the utility that is gained or lost as a result of the decisions taken with the aid of the algorithm. 

This general definition can be adapted to different contexts: In some contexts, what people desire can be measured in monetary terms. In other contexts, we may measure it on different scales, e.g., as health outcomes.
We rely on competent experts and stakeholders to identify a suitable operationalization of the concept of utility into something measurable, which is the second value-laden choice of our framework.

\subsection{Relevant groups}
\label{ssec:relevant_groups}
\textit{Between whom is it distributed?}

Most contemporary theories of justice focus on individuals, understood as bearers of utility, capabilities, or rights~\cite{sen1980equality}. Theories of discrimination, instead, relate to socially salient groups \cite{sep-discrimination}. We focus on a conception of "relevant groups", placing causal constraints on what qualifies \textit{a group} in a way that is relevant to group fairness. In our framework, relevant groups are defined by a \textit{weak causal link} in the context of the prediction-based decision in question.

\begin{definition}[Relevant groups]
Relevant groups are types of individuals that are representative of plausible causes of inequality in the outcome or in the prediction in the context to which the question of fairness relates.
\end{definition}

By invoking groups that satisfy a weak causal link we aim to address the objections typically raised against group fairness that we already mentioned in Section \ref{sec:related-work}, namely fairness gerrymandering and causal irrelevance. Proponents of individual fairness object that group fairness criteria are vulnerable to fairness gerrymandering: Any group fairness criterion can be satisfied by altering who receives a positive and negative decision, without improving the fairness of the treatment of any individual in that group~\cite{dwork2012fairness,raez2021group,baumann2022sufficiency}. Proponents of causal definitions of fairness object that group fairness criteria are unable to distinguish the case in which individuals of a group receive a worse outcome \textit{because} they are members of the group from those cases in which receiving a worse outcome is simply \textit{correlated} to being a member of the group, but the group does not as such \textit{influence} the decision.
Our weak causality requirement demands to only consider groups defined by features that are \textit{plausible} causal influences of the prediction or the outcome (or both), where causation can be both direct and indirect.
So, for example, in a racist society, race may define relevant groups; in a sexist society, gender may define relevant groups. The Cartesian product of the two (each combination of a race and gender variable) will then also be weakly causally relevant. Unlike counterfactual fairness, which requires modeling causal links between the prediction variable and the group variable, our framework takes the shortcut of only considering groups defined by features for which some degree of causal influence on the decision-relevant variable or the decision is plausible \textit{a-priori}, given what we know about society.
In principle, these groups \textit{could} be defined to be narrower and narrower.
This corresponds to the concept of multicalibration, which considers every efficiently-identifiable subgroup, i.e., the ``collection of subsets where set membership can be determined efficiently --- for instance, subpopulations defined by the conjunctions of a small number of boolean features or by small decision trees''~\cite[p.~1940]{hebert-johnson18Multicalibration}.
However, inequalities between very large numbers of extremely fine-grained groups are hard to morally judge in practice.
Therefore, in contrast to multicalibration, we consider only groups for which the weak causality requirement is satisfied instead of constraining the number of considered subgroups purely based on computational efficiency.
This means that our framework aims to compare groups (including those with very few and very similar individuals) identified by all the features that causally influence (directly or indirectly) the outcome or the prediction.
In most concrete contexts, a value-laden choice must be made to focus on one, or a few (intersectional) traits, guided by the concrete political priorities emerging in the context of our decision-making system and regarded most relevant by the stakeholders (step 3 of our framework).
Admittedly, this offers no guarantee that every observed inequality in average outcomes between groups is fully causally explained by the membership in those groups, so the approach is still vulnerable to counterexamples.
However, our conjecture is that the causal requirement makes it harder, in practice, to gerrymander fairness by characterizing inclusion and exclusion criteria of groups in an arbitrary manner (just for the sake of equalizing group frequencies) and it will raise the chances that the observed group inequality is --- to some degree if not entirely --- due to the groups being what they are.
We offer this as a mere empirical conjecture and as a pragmatic solution of the fairness gerrymandering and causal irrelevance problem.
It is a "solution" in a very different sense than the rigorous (formal) solutions offered by individual fairness~\cite{dwork2012fairness} and causal views of fairness~\cite{kusner2017counterfactual}.
As a practical method, our framework makes approximate fairness easier to achieve practice, because it has lower epistemic requirements than competing approaches, such as individual and counterfactual fairness.\footnote{Specifically, individual fairness can only be measured relative to an already given metric of similarity of individuals in the respect that matters to fairness. Clearly, unless the metric is itself objectively fair in a morally relevant sense, individually fair predictors cannot be considered fairer in a substantive sense than the predictors they aim to improve upon. For this, see~\cite{binns2020apparent} showing that this is in all but artificially defined cases extremely hard to satisfy.
Counterfactual fairness, on the other hand, can only be defined relative to an already given set of assumptions about the causal structure responsible for outcomes and decisions (e.g., a set of differential equations describing the direct and indirect influence of group features on both $Y$ and $D$).
In practice, it is extremely hard to know, justifiably believe, or even merely inter-subjectively agree upon a metric of similarity and a causal structure in the domains in which standard problems of fair AI emerge.} 

\subsection{Claim differentiator}

\textit{Which subgroups should be compared?}

In answering the question, "between whom are benefits and harms distributed", we must consider the following complications. In some contexts, comparisons between groups (defined by causal relevance) are not intuitively appropriate for fairness. This is because, in some cases, individuals within those groups who are different in some (morally salient) features should not be treated equally. According to contextualism\cite{miller1999principles}, what these morally salient features are depends on the context. For example, there are contexts in which individuals ought to be treated differently when their needs differ, but in other contexts individuals ought to be treated differently when their contributions differ. Moreover, in practice, we must deal with moral disagreement about whether need, responsibility, or contribution, for example, ought to matter, in a given context. To account for both contextual relativity and the possibility that not all stakeholder groups will adopt the same view of justified inequalities, we need to introduce a new parameter in the theory. This is the \textit{claim differentiator}, the feature(s) by virtue of which individuals can morally demand equal consideration of the harms and benefits produced by the algorithms. In practice, this parameter specifies the subgroups among the previously defined \textit{relevant groups} that should be compared.
This is the fourth value-laden choice of our framework.

\begin{definition}[Claim differentiator] A claim differentiator is a feature that distinguishes individuals who have different moral claims to utility. 
\end{definition}

We assume that the higher-order concept of a claim differentiator can be chosen on the basis of either the context or the moral theory endorsed by stakeholders. We introduce the claim differentiator as a novel concept that is not an established notion in political philosophy or moral philosophy.\footnote{A \textit{similar} idea is found in~\cite{loi2021fair,baumann2022SDS_fairness_principle,holm2022fairness}.} To provide more clarity about this higher-order moral concept, we provide in Section~\ref{ssec:Combining_relevant_groups_and_claims_differentiators} an analysis of luck-egalitarian equality of opportunity as the combination of substantive conception of the claim differentiator and a substantive conception of the pattern of justice, the element we introduce next. Moreover, we shall provide an example that illustrates the reasoning for the claim differentiator in a hypothetical concrete business scenario in Section \ref{ssec:Claims-differentiator-example}.

\subsection{Pattern of justice}
\textit{How should it be distributed?}

After discussing which groups have equal moral claims to the utility derived from the decisions, we have to consider whether we can tolerate inequalities in some cases.
One may say that inequalities are always unacceptable and that equality has to be achieved at all costs. However, this might result in leveling down: Assume a situation in which the utilities derived for the groups are unequal, but in order to equalize them, the utility of all groups has to be lowered. In that case, one might prefer the original unequal utility distribution, from which all groups profit. This is a well-known issue with existing group fairness metrics (see, e.g., \cite{hu2020fair, binns2018fairness,  feder2021emergent}). To avoid this, we can allow for some inequalities, e.g., if they are beneficial to the worst-off group.
Therefore, the fifth value-laden choice is to define what a just distribution looks like. This can be described as a \textit{pattern of justice}.

\begin{definition}[Pattern of justice]
A pattern of justice describes how utility should be distributed between the relevant groups.
\end{definition}

The most widely discussed patterns of justice in political philosophy are:
\begin{itemize}
    \item Egalitarianism \cite{sep-egalitarianism}: The group utility levels should be as equal as possible.
    \item Maximin \cite{rawls1999theory, rawls2001justice}: The goal is to maximize utility for the worst-off group.
    \item Prioritarianism \cite{holtug2017prioritarianism}: The goal is to maximize aggregate utility for all groups, giving greater weight to utility, the worse off the group.\footnote{Maximin is the extreme version of this as an infinite weight is given to the worst-off group.}
    \item Sufficientarianism \cite{shields2020sufficientarianism}: The goal is to bring all groups above a certain level of utility.
\end{itemize}

This also implies that the patterns have a different relationship to equality. Egalitarianism values equality above all else while the other patterns tolerate inequalities: Maximin tolerates inequalities if they profit the worst-off group; prioritarianism tolerates inequalities if they increase the aggregated utility; sufficientarianism tolerates inequalities as long as all groups achieve a minimum level of utility.

\subsection{Combining relevant groups, claim differentiators and pattern of justice}

\subsubsection{Combining relevant groups and claim differentiators}
\label{ssec:Combining_relevant_groups_and_claims_differentiators}

\begin{figure}[!htb]
    \centering
    \includegraphics[width=0.5\textwidth]{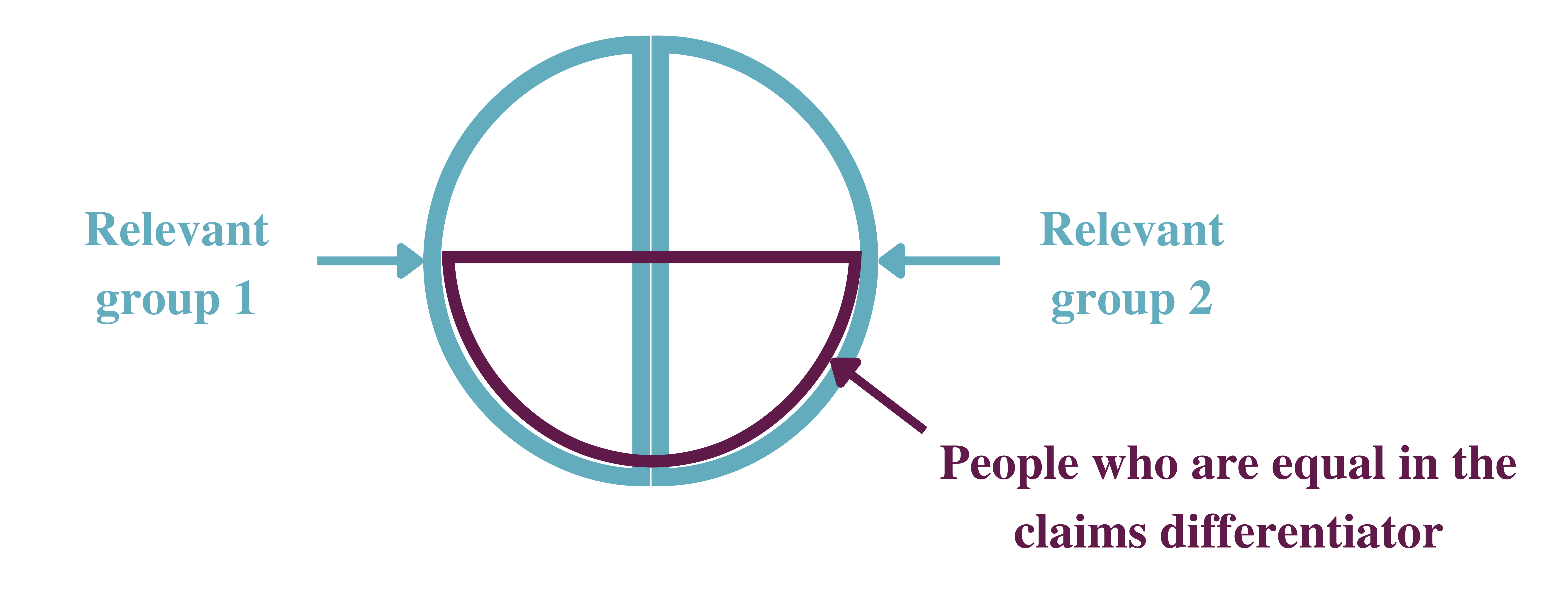}
    \caption{The relationship of relevant groups and the claim differentiator.}
    \label{fig:groups}
\end{figure}

Both concepts of relevant groups and the claim differentiator split the general population into subgroups. We can connect the two concepts by asking: Should we compare the relevant groups or only subgroups of the relevant groups? This is relevant as the relevant groups are, of course, not homogeneous but consist of many different individuals who may have different claims to utility. Figure \ref{fig:groups} shows this intuition of the claim differentiator as the selection of subgroups within the relevant groups. When we combine these two concepts, this gives us the groups whose utilities are to be compared from a fairness perspective: We analyze the distribution of utility between relevant groups, restricted to those individuals with the same moral claims (specified by the claim differentiator).

We provide a concrete example of the combination of relevant groups and the claim differentiator in Sections~\ref{ssec:Relevant-groups-example} and~\ref{ssec:Claims-differentiator-example}.

\subsubsection{Combining patterns of justice and claim differentiators}

We can view many influential theories of justice through the lens of these three components.
Egalitarian notions of fairness, for example, demand that people are equal in some regard \cite{sep-egalitarianism}.
Luck-egalitarian equality of opportunity can be defined as the view that individuals who make similar choices should have the same expectations or outcomes \cite{sep-equal-opportunity}. This is different from strict egalitarianism, which demands equality in outcomes \cite{sep-justice-distributive}. They differ in the claim differentiator: inequalities due to choices, not circumstances, are considered justified\cite{arneson1989equality, roemer1998equality}. Luck egalitarian equality of opportunity thus uses choices for which individuals are responsible as a claim differentiator.

It may be tempting to assume that the aforementioned definition of the claim differentiator implies that justice requires some kind of equality because individuals with the same value of the claim differentiator have morally the same claims to utility. However, this is wrong. Consider, for example, the combination of desert as a claim differentiator and maxmin as a pattern of justice. Justice is then achieved by maximizing the expectations of the worst-off relevant group \textit{among individuals who are equal in their contributions}. This is not a morally absurd view. For example, one may object to achieving equality between equal contributors (while recognizing that this is what they ideally deserve) when this, in the given circumstances, can only be achieved by leveling down.

\section{Trade-off between the decision maker's goals and fairness}\label{sec:pareto-front}

The previous steps allow us to define a fairness score, which quantifies the fairness of a decision rule, in a way that encapsulates the value choices listed above.\footnote{The mathematical details of how exactly to derive a fairness score from these fairness components is described in full detail in \cite{Baumann2023unification}.} Given a fairness score and a measure of expected decision-maker utility, it is possible to represent trade-offs in a bidimensional Cartesian plot \cite{kearns2019ethical}. It is reasonable to focus on the Pareto-efficient decision rules: those for which an improvement on one dimension is only possible if the other dimension is worsened. The last stage of moral discussion ought to concern the choice between points in the Pareto front, where any gain of fairness can only be achieved at the expense of the decision maker's utility, and vice-versa.

\section{Credit lending example}\label{sec:examples}

Let us now discuss a highly simplified example of financial lending to see how the perspectives of the decision maker (Section~\ref{sec:setting}) and the decision subjects (Section~\ref{sec:distributive-justice}) can be defined and balanced in practice.
Consider a bank's decision to accept a loan application ($D=1$) or reject it ($D=0$) based on scores representing repayment probabilities.
If the bank wants to balance its profitability with fairness, the first step is to specify how they measure its own utility, i.e., its profits. The subsequent steps of our framework determine how they measure fairness.
To illustrate this example, we are using the preprocessed version of the UCI German credit dataset and train a logistic regression to predict whether an individual will pay back a granted loan \cite{friedler2019comparative}.\footnote{
The code for this is publicly available at \url{https://github.com/hcorinna/utility-based-fairness}.
}

\subsection{Utility of the decision maker}

We start by specifying the perspective of the decision maker. Assuming that the bank is interested in profits, it has to assess how much profit is derived from each decision. As this illustration does not aim to be realistic, we do not consider the costs of the bank and assume that the interest payments\footnote{In the implementation of this example, we assume the interest rate to be 10\% for every loan regardless of risk or other factors.} are the profit of the bank while the cost of a defaulted loan is equivalent to the loan size. Rejected loan applications are defined as cost-neutral as we assume that the cost of reviewing applications is 0.
For a loan applicant $i$ with a repayment probability of $p_i$ asking for a loan of size $s_i$ with an interest rate of $z_i$, the bank's expected utility is thus $E(u_{DM,i}) = p_i \cdot  z_i  \cdot s_i - (1-p_i) \cdot s_i$. Utility-maximizing banks would grant a loan to all individuals with a positive expected utility (i.e., $D=1$ if $E(u_{DM,i})>0$).

\subsection{Utility of the decision subjects}

Next, the bank turns to the evaluation of how fair a given decision-making system is towards the decision subjects. For this, they might ask representatives of the decision subjects or experts from social sciences and philosophy to consider the components of fair decision-making described in Section~\ref{sec:distributive-justice}.

These representatives first have to answer the question of how to assess the utility that decision subjects derive from the decision-making process.
In the case of lending, \textit{loans} are distributed. Individuals do not profit equally from being granted a loan. If they cannot repay the loan and end up defaulting, it harms their future chances of receiving a loan.
To keep this example simple,\footnote{Additionally, we may consider factors such as the loan size or any other measurable attributes (e.g., individuals of a marginalized group might profit more from receiving a loan than individuals of a group that is better-off in many aspects of life).} we assume that stakeholders will base the utility assessment on the decision $D$ and whether the individual repays the granted loan $Y$. In our case, $D$ and $Y$ are binary variables, so there are four combinations whose utility we have to determine.

The utilities of the decision subjects can be visualized as a 2x2 matrix. Utility weights can be elicited through a dialogue between relevant stakeholders and experts on the impact of financial decisions. We note that, as long as there is a well-specified reference point and a corresponding scaling factor~\cite{elkan2001}, it is not necessary to express utility weights in terms of an external dimension (e.g., money) - weights matter only in so far as they define how morally beneficial or harmful a consequence is \textit{in proportion} to a different one.
For example:
\begin{itemize}
    \item $u_{D=1,Y=1}$: This asks for the utility of an individual who is granted a loan and repays it. Clearly, the individual derives a benefit from this: They receive the loan they applied for and can use it as planned. We assign a utility of +10.
    \item $u_{D=1,Y=0}$: This asks for the utility of an individual who is granted a loan and defaults. As stated above, the individual derives a harm from this: They receive the loan they applied for, but end up in debt as they cannot repay it. We assign a utility of -5.
    \item $u_{D=0,Y=1}$: This asks for the utility of an individual who is not granted a loan even though they would have been able to repay it. Their situation does not change much compared to their current situation. They have to invest additional time to apply for another loan, but assuming that there are other banks who will approve their loan application, this is only a small harm. We therefore assign a utility of -1.
    \item $u_{D=0,Y=0}$: This asks for the utility of an individual who is not granted a loan and would not have been able to repay it. Their situation does not change much compared to their current situation and given that they would not be able to repay their loan, they do not miss an opportunity by not being granted the loan. We therefore consider this combination to be neutral and assign a utility of 0.
\end{itemize}
If the property $u_{D=0,Y=0} \neq u_{D=1,Y=1}$ holds, we can to fix $u_{D=0,Y=0}'=0$ and $u_{D=1,Y=1}'=1$ so that the remaining utility weights can be expressed relative to this reference point and scale: $u_{D=1,Y=0}'=(u_{D=1,Y=0}-u_{D=0,Y=0})/(u_{D=1,Y=1}-u_{D=0,Y=0})$ and $u_{D=0,Y=1}'=(u_{D=0,Y=1}-u_{D=0,Y=0})/(u_{D=1,Y=1}-u_{D=0,Y=0})$.
This results in the utility matrix visualized in Table~\ref{tab:DS_utility_matrix}.
Notice that this shifting and scaling of all entries of the utility matrix does not affect the final decisions.
In practice, it is crucial to always elicit utility weights relative to a well-specified baseline.

\begin{table}[ht]
\caption{2x2 matrix representing the utility of the decision subjects}
\begin{tabular}{ccc}
                         & Y=0                     & Y=1                     \\ \cline{2-3} 
\multicolumn{1}{c|}{D=0} & \multicolumn{1}{c|}{0}  & \multicolumn{1}{c|}{-1} \\ \cline{2-3} 
\multicolumn{1}{c|}{D=1} & \multicolumn{1}{c|}{-5} & \multicolumn{1}{c|}{+10} \\ \cline{2-3} 
\end{tabular}
\label{tab:DS_utility_matrix}
\end{table}

\subsection{Relevant groups}\label{ssec:Relevant-groups-example}

The representatives next have to define the relevant groups to compare
and agree that groups defined by the sex attribute have unjustly unequal chances in life.\footnote{Even though sex is not binary, it is represented as a binary variable in this dataset.}
They therefore determine that the relevant groups to compare are women and men.

\subsection{Claim differentiator}\label{ssec:Claims-differentiator-example}
To determine the claim differentiator, the representatives have to answer the question "What makes it the case that \textit{certain individual types} (groups of people) have roughly the same claims to utility?" Suppose that the representatives agree that loan defaulters and non-defaulters cannot demand equal consideration. The only clients who have a claim to benefit from the decisions are those who will repay their debt. Therefore, they will compare the utility of people who repay their loan ($Y=1$).

\subsection{Pattern of justice}

We suppose that after deliberation, the representatives agree on the maximin pattern, so that the fairness score increases as the utility of the worst-off group increases.
When the bank and representatives compare different decision rules, they have to analyze how well these decision rules do with respect to maximin among the non-defaulters. This requires computing the expected utilities for both male and female non-defaulters under each decision rule and then comparing the lowest expected utilities.

\subsection{Trade-off decision}

From steps (2) to (5), it follows that the representatives decided to maximize the utility of the worst-off group between women and men who repay their loans ($Y=1$). However, suppose that this fairness goal conflicts with the decision maker's utility defined in step (1).
The last step in our framework is therefore to look at the trade-off between the goals of the decision maker and the fairness towards decision subjects.
As described in Section \ref{sec:pareto-front}, we use a Pareto front to visualize this trade-off for many different decision rules.
In line with~\cite{corbett2017algorithmic,baumann2022sufficiency}, we will assume that the decision rule takes the form of a threshold.\footnote{\cite{corbett2017algorithmic,baumann2022sufficiency} have proven this for egalitarian fairness criteria. We leave the proof that this also holds for maximin to future work.}

In this example, we test upper- and lower-bound thresholds for each group (men and women), resulting in the $(2*101)^2$ points seen in the Pareto plot in Figure~\ref{fig:pareto-front-all} (in the Appendix), where the Pareto front is marked in blue.\footnote{In principle, the number of thresholds that can be used for each group is infinite. In practice, we may plot the Pareto front for a very large number of thresholds combinations.}
The y-axis shows the decision maker's average utility per customer. A utility of, e.g., 10, means that the bank can expect a utility of 10 Deutsche Mark per customer. The x-axis shows the fairness score, which is the lower utility between the utility of women with $Y=1$ and men with $Y=1$.

Of course, we cannot claim that this Pareto plot shows the entire Pareto frontier as more points could be added. However, it visualizes some representative elements in the trade-off.

\begin{figure}[!htb]
    \centering
    \includegraphics[width=0.75\textwidth]{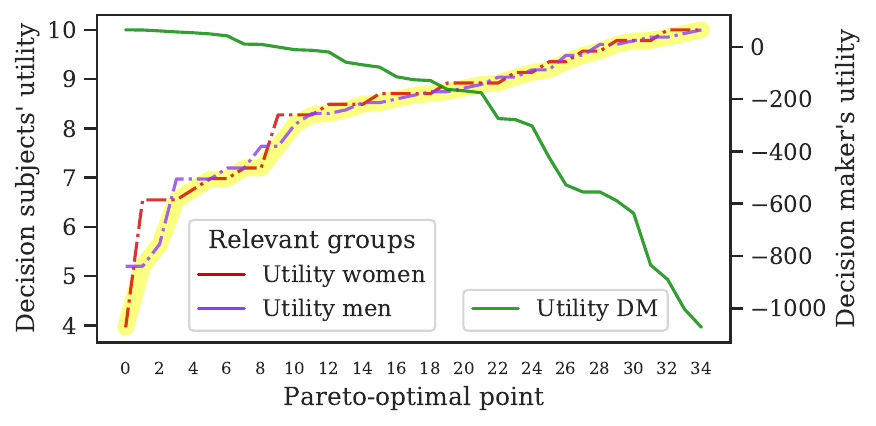}
    \caption{The DS utilities of women and men resulting from the decision rules on the Pareto front. The minimum expected DS utility (i.e., the fairness score for maximin) is highlighted in yellow.}
    \label{fig:utilities}
\end{figure}

As a start, the stakeholders, i.e., the bank and the representatives, may look at the two extreme points: the one that maximizes the utility of the decision maker (point 0) and the one that maximizes the fairness score (point 34). As can be seen in Figure~\ref{fig:utilities}, maximizing the utility of the decision maker results leads to inequality in the utility of women and men where women have the lower utility. With increasing fairness, the utility of the worst-off group (marked in yellow) also continuously increases. However, the difference in women's and men's utility does not continuously decrease even though the average utilities end up converging to the maximum possible expected value of 10 (which is achieved when everyone who is able to repay their loan receives a loan) for the fairest point (point 34).

\begin{figure}[!htb]
    \centering
    \includegraphics[trim={0 81.5cm 0 0}, width=0.9\textwidth]{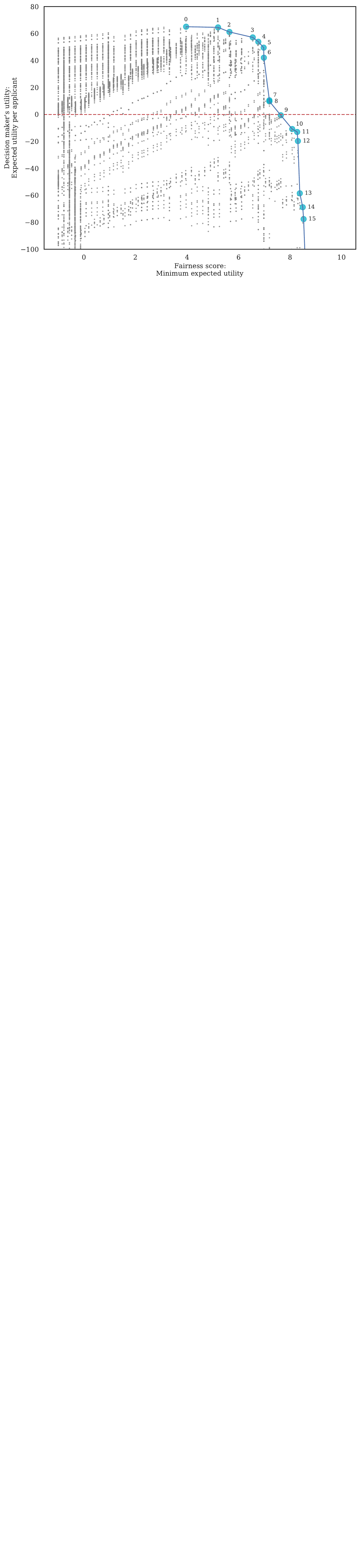}
    \caption{The Pareto front where the small, gray points are Pareto-dominated by the larger, blue points. Zoomed in to focus on the decision rules that are profitable for the bank.}
    \label{fig:pareto-front-zoom}
\end{figure}

Figures~\ref{fig:pareto-front-all} (in the Appendix) and~\ref{fig:utilities} also show the other points on the Pareto front and the corresponding utility values for women and men. As can be seen, most points on this Pareto front would lead to a negative expected utility for the bank (points below the red line in Figures~\ref{fig:pareto-front-zoom} and ~\ref{fig:pareto-front-all}). The bank does, of course, not consider such decision rules as it would sooner or later go out of business. Figure \ref{fig:pareto-front-zoom} therefore focuses on the profitable decision rules. Among those, the stakeholders may see points 3-5 as good trade-offs: They achieve a high fairness score with a high expected utility for both men and women while still being profitable for the bank. From this point on, one has to sacrifice a lot of the fairness score in order to gain a little in the utility of the decision maker (point 2), so the representatives may argue that this gain in the utility of the decision maker is too costly in terms of fairness.

It is important to note that our framework offers no principled solution to the problem of determining a valid trade-off value. The last step, unlike the previous five, is not guided by any kind of theory but it expresses the actual degree of attraction to fairness and aversion to loss of utility for the stakeholders consulted for this decision. The value of the framework is to lead the stakeholders to discuss \textit{necessary} sacrifices of utility or fairness, avoiding the selection of decision rules that are not Pareto optimal (given the value assumptions that have been - we suppose - antecedently agreed upon). 

\section{Limitations}\label{sec:limitations}

Interpersonal comparisons of well-being are notoriously difficult and here we rely on an objectivist view of well-being (which has to be elicited from experts) that may not correspond to the preferences and beliefs of the people affected by the decisions.
Furthermore, this approach is welfarist throughout. Welfarism is criticized by proponents of the capability approach as being too subjective \cite{sen1985standard}. Notice, however, that we do not rely on a preference-based or pleasured-based account of well-being, so we can include benefits to individual autonomy and freedom into our utility metric if stakeholders can agree on a suitable measurement. An objective measure of well-being can also consider the impact of resources on real individual freedoms (also known as capabilities) \cite{lundgard2020measuring}.\footnote{In this regard, we register disagreement between those, like \citet{sen1997choice}, who maintain that the value of individual agency cannot be captured by a welfarist framework and those, like \citet{Griffin1986WellBeing} who maintain that it is an element of (objective) well-being.}

Moreover, while our framework is compatible with many theories of distributive justice, it is not compatible with theories that do not follow the patterns of justice described in Section \ref{sec:distributive-justice}. This is, for example, the case for Nozick's entitlement theory \cite{nozick1974anarchy}.

On a practical level, it is not obvious how to make the six value-laden choices in practice and we only provided a sketch for this. More work needs to be done to deliver a practical empirical methodology to elicit the relevant value-laden choices from stakeholders.

This is perhaps why current group fairness metrics are so tempting: They do not require us to think through the choices of our framework. However, we must not delude ourselves: Not specifying every value-laden choice in our framework does not mean that we remain agnostic about what an appropriate choice might be --- we simply choose it implicitly. We argue that it is preferable to make these value-laden choices explicit in the design process. 

\section{Conclusion}\label{sec:conclusion}

With the increasing use of automated decision-making, there is a rising need to develop these systems ethically. This is not just a technical question but a question of values. However, these values are typically hidden in the technical details of the implementation. Like others before us, we therefore advocate for a more public debate about the values implemented in decision-making systems.

In this paper, we offer a framework to reveal and specify six key value-laden decisions behind the implementation of prediction-based decision-making systems. This includes the choice of a fairness criterion and the degree to which it can be enforced compatibly with the decision makers' original intentions.

Our framework helps bring the debate about values to the forefront of implementation rather than leaving these values as an accidental by-product. While our framework models more complex moral options than most, we kept it simple enough to be usable for actual stakeholder debates. We developed a web application that supports this deliberation process by visualizing fairness scores and their relation to the decision maker's utility.\footnote{\url{https://github.com/joebaumann/FairnessLab}}

\FloatBarrier

\begin{acks}
We thank the other members of our project and colleagues (Eleonora Viganò, Ulrich Leicht-Deobald, Serhiy Kandul, Markus Christen, Anikó Hannák, Nicolò Pagan, Stefania Ionescu, Aleksandra Urman, Leonore Röseler, Azza Bouleimen, and Egwuchukwu Ani) for their continuous feedback on the framework presented in this paper.
We also thank participants of our algorithmic fairness workshop at the Applied Machine Learning Days (AMLD) at École polytechnique fédérale de Lausanne (EPFL) in Switzerland and the participants of the course ``Informatics, Ethics and Society'' at the University of Zurich for critical discussions.
This work was supported by the National Research Programme “Digital Transformation” (NRP 77) of the Swiss National Science Foundation (SNSF) --- grant number 187473 --- and by Innosuisse --- grant number 44692.1 IP-SBM. Michele Loi was supported by the European Union's Horizon 2020 research and innovation programme under the Marie Sklodowska-Curie grant agreement No 898322.
\end{acks}

\bibliographystyle{ACM-Reference-Format}
\bibliography{main}

\appendix

\section{Additional Visualization}

\begin{figure}[!h]
    \centering
    \includegraphics[width=1\linewidth]{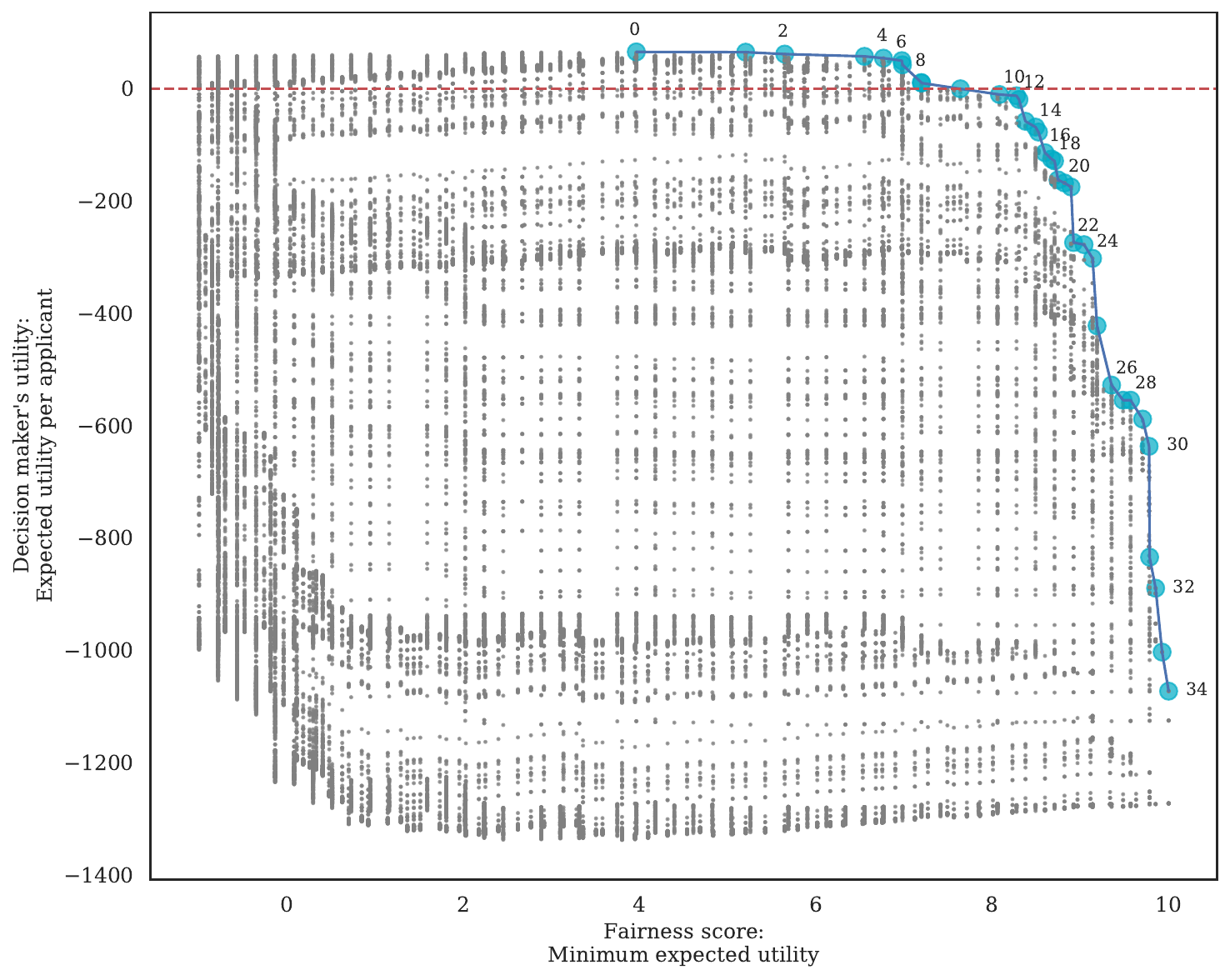}
    \caption{The full Pareto front where the small, gray points are Pareto-dominated by the larger, blue points.}
    \vspace{-20mm}
    \label{fig:pareto-front-all}
\end{figure}

\end{document}